\documentclass{article}
\usepackage{graphicx}
\usepackage{epsfig}
\usepackage{amsfonts}
\usepackage{slashed}

 \usepackage{amssymb}
\usepackage{amsmath}

\usepackage{hyperref}
\usepackage{graphicx}       
\usepackage{dcolumn}        
\usepackage{bm}             
\usepackage{amssymb}
\usepackage{amstext}
\usepackage{tensor}
\usepackage{cite}

 \usepackage{amssymb}

\date{\today}

\newcommand{\insertplot}[5]{\begin{figure}
 \hfill\hbox to 0.05in{\vbox to #5in{\vfill
 \inputplot{#1}{#4}{#5}}\hfill}
 \hfill\vspace{-.1in}
 \caption{#2}\label{#3}
 \end{figure}}
 \newcommand{\inputplot}[3]{
 \special{ps: plotfile #1}
}
\newcounter{fig}

\newcommand{\ee}{\end{equation}}
\newcommand{\eea}{\end{eqnarray}}
\newcommand{\be}{\begin{equation}}
\newcommand{\bea}{\begin{eqnarray}}

\begin{document}
\title{\Large{\bf Asymptotically flat spinning scalar, Dirac and Proca stars}}
 \vspace{1.5truecm}

\author{
{\large }
{\  C. Herdeiro}$^{1}$,
{\ I. Perapechka}$^{2}$,
{  E. Radu}$^{3}$,
and
{ Ya. Shnir}$^{4}$
\\
\\
$^{1}${\small CENTRA,  Departamento  de  F\'\i sica,  Instituto  Superior  T\'ecnico  -  IST,
}
\\
{\small
Universidade  de  Lisboa  -  UL,  Avenida  Rovisco  Pais  1,  1049  Lisboa,  Portugal}
\\
$^{2}${\small Department of Theoretical Physics and Astrophysics, Belarusian State
University,
}
\\
{\small
Nezavisimosti Avenue 4, Minsk 220004, Belarus}
\\
$^{3}${\small Departamento de F\'isica da Universidade de Aveiro and CIDMA,
}
\\
{\small
 Campus de Santiago, 3810-183 Aveiro, Portugal}
\\
$^{4}${\small
BLTP, JINR,
Joliot-Curie 6, Dubna 141980, Moscow Region, Russia}
}

\date{June 2019}

\maketitle

\begin{abstract}
Einstein's gravity minimally coupled to free, massive, classical fundamental fields admits particle-like solutions. These are asymptotically flat, everywhere non-singular configurations that realise Wheeler's concept of a \textit{geon}:
a localised lump of self-gravitating energy whose existence is anchored on the non-linearities of general relativity,  trivialising in the flat spacetime limit.  In~\cite{Herdeiro:2017fhv} the key properties for the existence of these solutions (also referred to as \textit{stars} or \textit{self-gravitating solitons}) were discussed -- which include a harmonic time dependence in the matter field --, and a comparative analysis of the stars arising in the Einstein-Klein-Gordon, Einstein-Dirac and Einstein-Proca models was performed, for the particular case of static, spherically symmetric spacetimes. In the present work we generalise this analysis for \textit{spinning} solutions. In particular, the spinning Einstein-Dirac stars are reported here for the first time.
Our analysis shows that the high degree of universality observed in the spherical case remains when angular momentum is allowed. Thus, as classical field theory solutions, these self-gravitating solitons are rather insensitive to the fundamental fermionic or bosonic nature of the corresponding field, displaying similar features. We describe some physical properties and, in particular, we observe that  the angular momentum of the
spinning stars satisfies the quantisation condition
$
J=m N,
$
for all models, where $N$ is the particle number and
 $m$ is an integer for the bosonic fields and a half-integer for the Dirac field. The way in which this quantisation condition arises, however, is more subtle for the non-zero spin fields.

\end{abstract}

\section{Introduction}
In vacuum Einstein's general relativity, the only physically reasonable stationary solution describing a
localised lump of energy is provided  by the Kerr black hole~\cite{Kerr:1963ud,Chrusciel:2012jk}.
A simple application of a Komar integral~\cite{Komar:1958wp} and the positive energy
theorem~\cite{Schon:1979rg,Witten:1981mf} shows that there are no everywhere regular localised lumps
of energy in vacuum, as realised (in a different way) long ago by Lichnerowicz \cite{Lichner}.

With some caveats (see, $e.g.$ the discussion in~\cite{Herdeiro:2019oqp}), the situation is similar
if Einstein's gravity is minimally coupled to a massless, free, fundamental field.
This includes, in particular, electrovacuum. But a rather distinct situation becomes possible
if the fundamental field is massive and with enough degrees of freedom. Considering a massive complex
Klein-Gordon, or Dirac or Proca field, minimally coupled to Einstein's gravity,  everywhere regular
localised solutions are possible - see \cite{Kaup:1968zz,Ruffini:1969qy,Finster:1998ws,Brito:2015pxa},
for the original references.\footnote{The inclusion of matter self-interactions opens
the possibility of  particle-like objects
with finite energy also in flat spacetime  - see \cite{Radu:2008pp} for a review -
albeit only bosonic such solutions have been so far considered.
 In this work we shall restrict ourselves to free matter fields.} We shall refer to these self-gravitating
 solitonic solutions as, respectively, scalar, Dirac or Proca stars, which
 provide explicit realisations of Wheeler's \textit{geons} \cite{Wheeler:1955zz}.
 Naturally, they where originally computed under the assumption of a spherically symmetric,
 static spacetime. Yet, rotation is ubiquitous, for all objects, in all scales. Thus, despite
 the higher technical complexity, it is of interest to study rotating scalar, Dirac or Proca stars.
 For the bosonic fields, the corresponding spinning stars were first computed
 in~\cite{Schunck:1996he,Yoshida:1997qf,Brito:2015pxa,Herdeiro:2016tmi,Herdeiro:2017phl}, whereas for the
 Dirac case they will be described herein for the first time. This is one of the main purposes of this work.

 It turns out that a spinning Dirac star is somewhat more natural than the static spinless one. Indeed, since
a single fermion possesses an intrinsic angular momentum, the matter content required to obtain a spinless
solution consists of  (at least)
two fermionic fields which allows for an angular momentum cancellation. To study spinning Dirac stars, on
the other hand we need a single Dirac field.  With respect to their bosonic counterparts, which can be
regarded as {\it `macroscopic quantum states'} prevented from gravitationally
collapsing by Heisenberg's uncertainty principle, the interpretation of the Dirac stars is more delicate
and has been considered in~\cite{Herdeiro:2017fhv}. As classical field theory solutions, however, Dirac stars are
in many ways similar to the bosonic ones, an observation already established in~\cite{Herdeiro:2017fhv} for
the static case and confirmed here for the spinning solutions. For instance,   rotating Dirac stars 
have an intrinsic toroidal topology in their energy distribution, which parallels that of the rotating scalar stars~\cite{Schunck:1996he}; for all cases, moreover, the star's angular momentum $J$ is quantised as
$
J=m Q,
$
where $m$ is an integer
and $Q$ the Noether charge, that also becomes an integer $Q=N$ upon quantisation.
To make this comparison more meaningful, following~\cite{Herdeiro:2017fhv}, we analyse the three types of stars
under a unified framework. Thus, the mathematical description of each of the
three models is made in parallel to emphasise the similarities.
The physical interpretation is only
distinct when quantisation is taken into account, which distinguishes fermions and bosons.
Then, in particular, whereas the bosonic configurations form
a continuous sequence or family of solutions for a given field mass,
 fermionic solutions do not, due to  Pauli's exclusion principle~\cite{Herdeiro:2017fhv}.

This paper is organised as follows. In Section 2
we describe the basic equations of each of the three different
models. In Section 3 we introduce the spacetime and matter fields ansatz. In Section 4 we
discuss the global quantities and the angular momentum-Noether charge relation which is universal
for the three models but appears in a more contrived way in the cases with non-zero spin. In Section 5
we  construct the spinning stars by solving
numerically the field equations subject to specified boundary conditions.
 We also clarify the physical interpretation of the sequences of fermionic solutions.
Concluding remarks and some open questions are presented in Section 6.

\section{The model}

Let us first describe the three models. The discussion and conventions follow closely
those in~\cite{Herdeiro:2017fhv} where a few more details are provided. Einstein's gravity in 3+1 dimensional spacetime is
minimally coupled with a spin-$s$ field, where $s$ takes one of the values $s=0,\frac{1}{2},1$.
The action is (with $c=1=\hbar$)
\begin{eqnarray}
\label{action}
\mathcal{S}=\int d^4 x \sqrt{-g}
\left [
\frac{R}{16 \pi G}
+
\mathcal{L}_{(s)}
\right] \ ,
\end{eqnarray}
where the three possible matter Lagrangians are:
\begin{eqnarray}
\label{LS}
&& \mathcal{L}_{(0)}= - g^{\alpha \beta}\bar \Phi_{, \, \alpha} \Phi_{, \, \beta} - \mu^2 \bar \Phi \Phi \ , \qquad \mathcal{L}_{(1)}= -\frac{1}{4}\mathcal{F}_{\alpha\beta}\bar{\mathcal{F}}^{\alpha\beta}
-\frac{\mu^2}{2}\mathcal{A}_\alpha\bar{\mathcal{A}}^\alpha \ ,
\\
\label{LD}
&&
 \mathcal{L}_{(1/2)} =-i
\left[
\frac{1}{2}
  \left( \{ \hat{\slashed D}  \overline{\Psi}   \} \Psi  -
     \overline{\Psi}  \hat{\slashed D}  \Psi
    \right)
+\mu \overline{\Psi}   \Psi
\right]\ .
\end{eqnarray}
Here, $\Phi$ is a complex scalar field;
$\Psi$ is a Dirac 4-spinor, with four complex components; 
 $\hat{\slashed D}\equiv \gamma^\mu \hat{D}_\mu$, where  $\gamma^\mu$ are the curved spacetime gamma matrices,
$\hat{D}_\mu=\partial_\mu - \Gamma_\mu$
is the spinor covariant derivative and $\Gamma_\mu$ are the spinor connection matrices~\cite{Dolan:2015eua};
$\mathcal{A}$ is a complex 4-potential, with the field strength $\mathcal{F} =d\mathcal{A}$.
In all cases, $\mu>0$ corresponds to the mass of the field(s).
For the scalar and Proca fields, the overbar denotes complex conjugation; $\overline{\Psi}$ denotes the Dirac conjugate \cite{Dolan:2015eua}.

\medskip

Variation of (\ref{action}) with respect to the metric leads to the Einstein field equations
\begin{eqnarray}
E_{\alpha \beta}=G_{\alpha \beta}-8 \pi G~ T_{\alpha \beta}^{(s)}=0\ ,
\end{eqnarray}
 where $G_{\alpha \beta}$ denotes, as usual, the Einstein tensor and
$T_{\alpha \beta}^{(s)}$
is the energy-momentum tensor:
\begin{eqnarray}
\label{TS}
&&
T_{\alpha \beta}^{(0)}=
\bar  \Phi_{ , \alpha}\Phi_{,\beta}
+\bar \Phi_{,\beta}\Phi_{,\alpha}
-g_{\alpha \beta}  \left[ \frac{1}{2} g^{\gamma \delta}
 ( \bar \Phi_{,\gamma}\Phi_{,\delta}+
\bar \Phi_{,\delta}\Phi_{,\gamma} )+\mu^2 \bar \Phi\Phi\right] \ ,
\\
&&
T_{\alpha \beta}^{(1/2)}= -\frac{i}{2}
\left[
    \overline{\Psi}  \gamma_{(\alpha} \hat{D}_{\beta)} \Psi
-  \left\{ \hat{D}_{(\alpha} \overline{\Psi} \right\} \, \gamma_{\beta)} \Psi
\right]  \ ,
\label{TD}
\\
&&
T_{\alpha\beta}^{(1)}=\frac{1}{2}
( \mathcal{F}_{\alpha \sigma }\bar{\mathcal{F}}_{\beta \gamma}
+\bar{\mathcal{F}}_{\alpha \sigma } \mathcal{F}_{\beta \gamma}
)g^{\sigma \gamma}
-\frac{1}{4}g_{\alpha\beta}\mathcal{F}_{\sigma\tau}\bar{\mathcal{F}}^{\sigma\tau}+\frac{\mu^2}{2}\left[
\mathcal{A}_{\alpha}\bar{\mathcal{A}}_{\beta}
+\bar{\mathcal{A}}_{\alpha}\mathcal{A}_{\beta}
-g_{\alpha\beta} \mathcal{A}_\sigma\bar{\mathcal{A}}^\sigma\right] \ .~{~~}
\label{TP}
\end{eqnarray}
The corresponding matter field equations are:
\begin{eqnarray}
&& \nabla^2 \Phi-\mu^2\Phi=0 \ ,
\qquad
\hat{\slashed D}\Psi    - \mu \Psi    = 0 \ ,
\qquad
\nabla_\alpha\mathcal{F}^{\alpha\beta}-\mu^2 \mathcal{A}^\beta=0\ .
\label{LP2}
\end{eqnarray}
In the Proca case,
the field eqs. (\ref{LP2}) imply the Lorentz condition,  $\nabla_\alpha\mathcal{A}^\alpha = 0$.

The matter field action, in all cases,
possesses a
  global $U(1)$ invariance, under the transformation
    $\{ \Phi, \Psi, \mathcal{A}\} \rightarrow e^{i a}\{ \Phi, \Psi, \mathcal{A}\} $,
    where $a$ is a constant.
By Noether's theorem this implies the existence of a conserved 4-current:
\begin{eqnarray}
\label{jS}
&& j^\alpha_{(0)}=-i (\bar \Phi \partial^\alpha \Phi-\Phi \partial^\alpha \bar \Phi) \ , \qquad  j^\alpha_{(1/2)}=\bar \Psi \gamma^\alpha \Psi \ , \qquad  j^\alpha_{(1)}=
\frac{i}{2}\left[\bar{\mathcal{F}}^{\alpha\beta}\mathcal{A}_\beta-\mathcal{F}^{\alpha \beta}\bar{\mathcal{A}}_\beta\right]\ .
\end{eqnarray}
Indeed, the field equations imply $j^{\alpha}_{ (s) ;\alpha}=0$. Then, integrating the timelike component of this 4-current on a spacelike hypersurface $\Sigma$ yields a conserved  \textit{Noether charge}:
\begin{eqnarray}
\label{Q}
Q_{(s)}=\int_{\Sigma}~j^t _{(s)}\ .
\end{eqnarray}
The Noether charge become an integer after quantisation, $Q=N$, where $N$ is the particle number.

\section{The ansatz}
\label{sec3}

We seek spacetimes with  two commuting Killing vector fields,
$\xi$ and $\eta$, with
$
\xi=\partial_t ,
$
and
$\eta=\partial_{\varphi}
$,
in a coordinate system adapted to the isometries, where $t$ and $\varphi$ are the time and azimuthal coordinates, respectively.
General relativity solutions with these symmetries are usually studied
within the following  metric ansatz:
$
ds^2=-e^{-2U(\rho,z)} (dt+\Omega(\rho,z)d \varphi)^2+e^{2U(\rho,z)}\left(e^{2k(\rho,z)}(d\rho^2+dz^2)+S^2(\rho,z)d\varphi^2 \right)
,
$
where $(\rho,z)$ correspond, asymptotically, to standard cylindrical coordinates.
 In the electrovacuum case, it is always possible
to set $S\equiv \rho$,
such that only three independent metric functions appear in the equations,
and $(\rho,z)$ become  the canonical Weyl coordinates \cite{book}.
For the matter sources in this work, however,
a generic metric ansatz with four independent functions is needed.
Also, it turns out to be more  convenient for numerics
to use `spheroidal-type' coordinates $(r,\theta)$
defined as
$
\rho=r\sin \theta,~z=r\cos \theta~,
$
instead  of $(\rho,z)$, with
 the usual range $0\leqslant r<\infty$, $0\leqslant \theta \leqslant \pi$.
After a suitable redefinition of the metric functions,
this leads to the  following metric ansatz:
\begin{eqnarray}
\label{metric}
ds^2=-e^{2F_0} dt^2+e^{2F_1}\left(dr^2+r^2 d\theta^2\right)+e^{2F_2}r^2 \sin^2\theta \left(d\varphi-\frac{W}{r} dt\right)^2\ ,
\end{eqnarray}
which has been employed in the study
of $s=0$
\cite{Herdeiro:2015gia}
and $s=1$
\cite{Brito:2015pxa,Herdeiro:2016tmi}
 spinning stars.
The four metric functions $(F_i;W)$, $i=0,1,2$, 
are functions of the variables $r$ and $\theta$ only, chosen such that
the trivial angular and radial dependence of the line element is already factorised.
The symmetry axis of the spacetime is given by $\eta^2=0$ and
 corresponds to $\theta=0,\pi$.
The Minkowski spacetime background is approached for $r\to \infty$, where the asymptotic values are $F_i=0$, $W=0$.

For the Dirac stars case ($s=1/2$),
we shall employ the following orthonormal tetrad for the metric (\ref{metric})
\begin{equation}
{\bf e}^0_\mu dx^\mu=e^{F_0}dt\ , \qquad {\bf e}^1_\mu dx^\mu=e^{F_1}dr\ , \qquad {\bf e}^2_\mu dx^\mu=e^{F_1}r d\theta\ ,
\qquad {\bf e}^3_\mu dx^\mu=e^{F_2}r \sin \theta \left(d\varphi-\frac{W}{r}dt\right) \ ,
\end{equation}
such that $ds^2=\eta_{ab}({\bf e}^a_\mu dx^\mu) ({\bf e}^b_\nu dx^\nu)$, where $\eta_{ab}={\rm diag}(-1,+1,+1,+1)$.

Let us now consider the ansatz for the three mater fields.  In the scalar case, the matter field ansatz which is compatible with
an axially symmetric geometry is written in terms of a single real function $\phi(r,\theta)$, and reads:
\begin{eqnarray}
\label{S}
&&
\Phi=e^{i(m \varphi-w t)}\phi(r,\theta) \ .
\end{eqnarray}
In the Proca case, the ansatz introduces four real potentials~\cite{Brito:2015pxa}:
\begin{eqnarray}
\label{P}
&&
\mathcal{A}=e^{i(m\varphi-w t)}\left(
 iV(r,\theta) dt  +\frac{H_1(r,\theta)}{r}dr+H_2(r,\theta)d\theta+i H_3(r,\theta) \sin \theta d\varphi
\right) \ .
\end{eqnarray}
In the case of a Dirac field, the ansatz
also contains four real functions\footnote{Ansatz (\ref{D}) is compatible with the
(circular) metric form (\ref{metric}).
Also, the ansatz
considered in \cite{Finster:1998ws,Herdeiro:2017fhv}
in the study of spherically symmetric stars is
recovered for $m=\pm 1/2$,
with a factorised angular dependence.}
\begin{eqnarray}
\label{D}
\Psi = e^{i(m\varphi-w t) } \begin{pmatrix}
\psi_1 (r,\theta)
\\
\psi_2 (r,\theta)
\\
-i \psi_1^* (r,\theta)
\\
-i\psi_2^* (r,\theta)
\end{pmatrix}
\ ,~~
{\rm with}~~\psi_1 (r,\theta)=P(r,\theta)+i Q(r,\theta)\ ,~~
\psi_2 (r,\theta)=X(r,\theta)+i Y(r,\theta)\ .
\end{eqnarray}
For $s=0,1$, the parameter $m$ in an integer, while for the Dirac field
  $m$ is a half-integer; $w$ is the field's frequency in all three cases, which we shall take to be positive.

\section{Global charges and the $J$-$Q$ relation}

Given the above ansatz, let us consider the explicit form for two relevant physical quantities.
The first one is
 the temporal component of the current density:
\begin{eqnarray}
\label{Q0}
&&
j^{t}_{(0)}=2e^{-2F_0} \left(w-\frac{m W}{r}\right) \phi^2 \ ,
\\
\label{Q12}
&&
j^{t}_{(1/2)}=2e^{-F_0}(P^2+Q^2+X^2+Y^2)\ ,
\\
\label{Q1}
 &&
j^{t}_{(1)}
= \frac{e^{-2(F_0+F_2)}}{r^2}H_3\left(wH_3+\frac{m V}{\sin \theta}\right)
+\frac{e^{-2(F_0+F_1)}}{r^3}
\bigg  \{
r(H_1^2+H_2^2)\left(w-\frac{m W}{r}\right) \nonumber
\\
&&{~~~~~}
  +\cos\theta H_2H_3 W
+r H_1 (rV_{,r}+\sin \theta W H_{3,r})
+H_2 (rV_{,\theta}+\sin \theta W H_{3,\theta})
\bigg  \}\ .
 \end{eqnarray}
The second one is the angular momentum density:
 \begin{eqnarray}
\label{J0}
&&T_\varphi^{t(0)}=
2e^{-2F_0}m  \left(w-  \frac{mW}{r}\right)\phi^2\ ,
\label{J12}
 \end{eqnarray}
 \begin{eqnarray}
&&
T_\varphi^{t(1/2)}=
e^{-F_0}m (P^2+Q^2+X^2+Y^2)
+
e^{-F_0-F_1+F_2}\sin \theta
\left\{
(PX+QY)[1+r(F_{2,r}-F_{0,r})] \nonumber \right.
\\
&&{~~~~~} \left.
-\frac{1}{2}
(P^2+Q^2-X^2-Y^2)
(\cot \theta+F_{2,\theta}-F_{0,\theta})
+2e^{-F_0+F_1} r \left(w-\frac{m W}{r}\right)(QX-PY)
\right\} \ , \ \
\\
\label{J1}
&&
T_\varphi^{t(1)}=
-\frac{\mu^2}{r}e^{-2F_0}
H_3 \sin \theta
(
rV+H_3 \sin \theta W
)
+\frac{e^{-2(F_0+F_1)}}{r}
\left\{
\frac{H_1}{r}
\sin \theta (-rw +2 m W)H_{3,r} \right. \nonumber
\\
\nonumber
&&{~~~~~}
+(mH_1-r \sin \theta H_{3,r})V_{,r}
-W [\sin^2\theta H_{3,r}^2+\frac{1}{r^2}(\cos\theta H_3+\sin \theta H_{3,\theta})^2 ]
+\frac{mH_2 V_{,\theta}}{r}
\\
&&{~~~~~} \left.
+\frac{1}{r^2}(\cos\theta H_3+\sin \theta H_{3,\theta})
[H_2(r w-2m W)+r V_{,\theta}]
+\frac{m}{r}(H_1^2+H_2^2)\left(w-\frac{m W}{r}\right)
\right\}~.
\end{eqnarray}

 The ADM mass $M$ and the angular momentum $J$ of the solutions are read off from the asymptotic expansion:
\begin{eqnarray}
\label{asym}
g_{tt} =-1+\frac{2M}{r}+\dots\ , \qquad g_{\varphi t}=-\frac{2J}{r}\sin^2\theta+\dots \ . \ \ \
\end{eqnarray}
The total angular momentum can also be computed as the integral of the corresponding density\footnote{The ADM mass can also be computed as volume integral; however, this is less relevant in the context of this work.}
\begin{equation}
\label{Js}
J \equiv J_{(s)} =2\pi \int^{\infty}_0 dr\, \int^{\infty}_0 d\theta r^2 e^{F_0+2F_1+F_2 }T_\varphi^{t(s)} \ .
\end{equation}
The explicit form of the Noether charge,
as computed from~\eqref{Q},  is
\begin{equation}
\label{Qs}
Q \equiv  Q_{(s)}= 2\pi \int^{\infty}_0 dr\, \int^{\infty}_0 d\theta r^2 e^{F_0+2F_1+F_2 }j^{t}_{(s)} \ .
\end{equation}
For a scalar field  one can easily see that $J$ and $Q$
are proportional,
\begin{equation}
\label{JQ}
J=m Q \ ,
\end{equation}
since the corresponding densities
(\ref{Q0}),
(\ref{J0}),
 are identical up to a factor of $m$.
It turns out that this relation also holds for the Dirac and Proca case, but the result is
 less obvious,
since
 the angular momentum \textit{density} and Noether charge \textit{density} are $not$ proportional. Nonetheless,
the proportionality still holds at the level of the integrated quantities. Indeed,  in both cases
the angular momentum density and Noether charge density (multiplied by the azimuthal index $m$)
differ by a total divergence,\footnote{In deriving (\ref{rel-dif}) one uses also the matter field equations.}
\begin{eqnarray}
\label{rel-dif}
T_\varphi^t =m j^t
+ \nabla_\alpha P^\alpha \ ,
\end{eqnarray}
with
\begin{eqnarray}
 P^\alpha= {\mathcal{A}}_\varphi \bar{ {\mathcal{F}}}^{\alpha t}+\bar{\mathcal{A}}_\varphi { {\mathcal{F}}}^{\alpha t} \ ,
\end{eqnarray}
for the Proca field \cite{Herdeiro:2016tmi}, and
\begin{eqnarray}
\label{s4}
  P^\alpha=  -\frac{i}{4}  \overline{\Psi}  \gamma_{\varphi } \gamma^{\alpha}\gamma^{  t} \Psi \ ,
\end{eqnarray}
for the Dirac field.
The total divergence is non-zero locally;
however, its volume integral vanishes for the solutions subject to the boundary conditions described in the next Section.
As a result, (\ref{JQ}) still holds for a Proca and Dirac fields.
Observe, nonetheless, the implicit differences in this relation. The bosonic solutions with $m=0$
are static; but for the Dirac stars $m$ is a half-integer and thus  cannot be zero -- they are necessarily rotating (recall static Proca stars require at least two Dirac fields).

The solutions satisfy also a first law of thermodynamics of the type:
\begin{eqnarray}
 dM=w dQ \ ,
\end{eqnarray}
which provides a test of numerical accuracy.

\section{The solutions}
\label{sec4}

In solving the equations of motion we exploit some symmetries thereof. Let us briefly comment on these, following \cite{Herdeiro:2017fhv}. Firstly, the factor of $4 \pi G$ in the Einstein field equations can be set to unity by a redefinition of
 the matter functions
\begin{equation}
\{\Phi,\mathcal{A},\Psi\}  \to \frac{1}{\sqrt{4\pi G}}  \{ \Phi,\mathcal{A},\Psi \} \ .
\label{t-D}
\end{equation}

Secondly, the field equations remain invariant under the transformation
\begin{equation}
\label{sca1}
(*):~~~
r \to \lambda r \ , \ \ \ W \to \lambda W \ , \,  \ \ \ F_i \to F_i \ ,\ \ \{w,\mu\} \to \frac{1}{\lambda}\{w,\mu\}\ ,
 \ \ \
\left\{
\begin{array}{l}
\displaystyle{\Phi \to \Phi } \ , \\
 \mathcal{A}  \to \frac{1}{\sqrt{\lambda}}\mathcal{A}  \ , \\
\Psi \to \Psi  \ .
\end{array}
\right \},
\end{equation}
where $\lambda$ is a positive constant.
In all three cases the ratio $w/\mu$ is left invariant by the $(*)$ symmetry.
This $(*)$-invariance is used to work in units set by the field mass,
\begin{eqnarray}
\bar \mu=1\ , ~~i.e.~~\lambda=\frac{1}{\mu}\ .
\label{mbar}
\end{eqnarray}
Then, to recover the physical quantities from those obtained in the numerical solution, a set of relations are used, identical to the ones described in~\cite{Herdeiro:2017fhv}.

\subsection{The boundary conditions and numerical method}

Given the matter  ansatz (\ref{S})-(\ref{D}), all components of the energy momentum tensor are  zero,
except for
$
T_{rr},~T_{r\theta},~T_{\varphi \varphi},~T_{t t}
$
and
$T_{\varphi t}$,
which possess a $(r,\theta)$-dependence only.
Then, 
the Einstein field equations with the energy momentum-tensors \eqref{TS}-\eqref{TP},
plus the matter field equations \eqref{LP2},
together with the ansatz \eqref{S}-\eqref{D},
lead to a system of five (eight) coupled partial differential equations for the  scalar (Dirac and Proca) cases.
There are four equations for the metric functions $F_i,W$; these are found by
taking suitable combinations of the Einstein equations:
$E_r^r+E_\theta^\theta=0$,
$E_\varphi^\varphi=0$,
$E_t^t=0$ and $E_\varphi^t=0$; 
additionally, there is one (four) equations for the matter functions.
Apart from these, there are two more Einstein equations
  $E_\theta^r =0,~E_r^r-E_\theta^\theta  =0$,
  which are not solved in practice.
Following an argument originally proposed in \cite{Wiseman:2002zc},
one can, however, show that
the identities $\nabla_\nu E^{\nu r} =0$ and $\nabla_\nu E^{\nu \theta}=0$,
imply the Cauchy-Riemann relations
$
\partial_{\bar r} {\cal P}_2  +
\partial_\theta {\cal P}_1
= 0 ,
$
$
 \partial_{\bar r} {\cal P}_1
-\partial_{\theta} {\cal P}_2
~= 0 ,
$
with ${\cal P}_1=\sqrt{-g} E^r_\theta$, ${\cal P}_2=\sqrt{-g}r(E^r_r-E^\theta_\theta)/2$
and $d\bar r=dr/{r }$.
Therefore the weighted constraints  $E_\theta^r$ and $E_r^r-E_\theta^\theta$
still satisfy Laplace equations in $(\bar r,\theta)$ variables.
Then they
are fulfilled, when one of them is satisfied on the boundary and the other
at a single point
\cite{Wiseman:2002zc}.
From the boundary  conditions below,
it turns out that this is the case for all three models,
 $i.e.$ the numerical scheme is self-consistent.

The boundary conditions are found by considering an approximate construction of the solutions
on the boundary of the domain of integration
together with the assumption of regularity and asymptotic flatness.\footnote{
In particular, the matter field equations in the far field reveal that
the solutions satisfy the condition $w<\mu$.}
The metric functions satisfy
\begin{equation}
\partial_r F_i\big|_{r=0}=W\big|_{r=0}=0 \ , \qquad
 F_i\big|_{r=\infty}=W\big|_{r=\infty}=0 \ , \qquad \partial_\theta F_i\big|_{\theta=0,\pi}=\partial_\theta W\big|_{\theta=0,\pi}=0 \ .
 \end{equation}
The scalar field amplitude vanishes
on the boundary of the domain of integration
 (see $e.g.$ \cite{Yoshida:1997qf})
\begin{equation}
\phi \big|_{r=0}= \phi \big|_{r=\infty}=\phi \big|_{\theta=0,\pi}=0 \ .
\end{equation}
The  boundary conditions in the  Proca case are
\cite{Brito:2015pxa,Herdeiro:2016tmi},
\begin{equation}
H_i|_{r=0}=V|_{r=0}=0\ , \qquad H_i|_{r=\infty}=V|_{r=\infty}=0 \ ,  \qquad H_1|_{\theta=0,\pi}=\partial_\theta H_2\big|_{\theta=0,\pi}=\partial_\theta H_3\big|_{\theta=0,\pi}=V|_{\theta=0,\pi}=0 \ ,
\end{equation}
where the last set of conditions applies to the lowest  $m=1$ states.
 For a Dirac field, one imposes
\begin{equation}
P\big |_{r=0}=   Q\big |_{r=0}= X\big |_{r=0}=  Y\big |_{r=0}=0 \ , \qquad
 P\big |_{r=\infty}=Q\big |_{r=\infty}=X\big |_{r=\infty}=Y\big |_{r=\infty}=0 \ ,
 \end{equation}
and,
for $m=1/2$,
\begin{equation}
\partial_\theta P\big |_{\theta=0 }= \partial_\theta Q\big |_{\theta=0 }=X\big |_{\theta=0 }=Y\big |_{\theta=0 }=0\ , \qquad
 P\big |_{\theta=\pi }=   Q\big |_{\theta= \pi}=\partial_\theta X\big |_{\theta=\pi }= \partial_\theta Y\big |_{\theta=\pi }=0 \ .
\end{equation}

In all three cases,
the solutions are found by
using a fourth order
finite difference scheme.
The system of five/eight equations is discretised on a grid with
$N_r\times N_\theta$ points
(where typically $N_r\sim 200$, $N_\theta \sim 50$).
We also introduce
a new radial coordinate $x={r}/({r+c})$, which maps the semi-infinite region $[0,\infty )$ onto the unit interval
$[0,1]$ (with $c$ some constant of order one).
The bosonic stars  were constructed by
using the professional package FIDISOL/CADSOL
 \cite{schoen}
 which uses a Newton-Raphson method.
The Einstein-Dirac system is solved with the Intel MKL PARDISO sparse direct solver \cite{pardiso},
and using the CESDSOL\footnote{Complex Equations -- Simple
Domain partial differential equations SOLver is a C++ package
being developed by one of us (I.P.).
} library.
In all cases, the typical
errors are of order of $10^{-4}$.

The data shown in this work correspond to fundamental states,
 all matter functions being nodeless.\footnote{
For a given $w$,
 a discrete set of solutions may exist, indexed by the number of nodes, $n$,
of (some of) the matter function(s).
Such excited solutions were reported for $s=0$
(see $e.g.$ \cite{Kunz:2019bhm})
and $s=1$ fields
(see
 \cite{Brito:2015pxa,Herdeiro:2016tmi}).
}
For the solutions herein,
the geometry  and the matter/current distributions are
invariant under a reflexion in the equatorial plane $(\theta=\pi/2)$, thus possessing a $\mathbb{Z}_2$ symmetry. 
Also, we shall consider solutions with the lowest number $m$
(except for the Dirac stars in Figure \ref{fig4}, right panel).

\subsection{Numerical results: basic properties
and domain of existence }

\begin{figure}[h!]
\begin{center}
\includegraphics[width=0.49\textwidth]{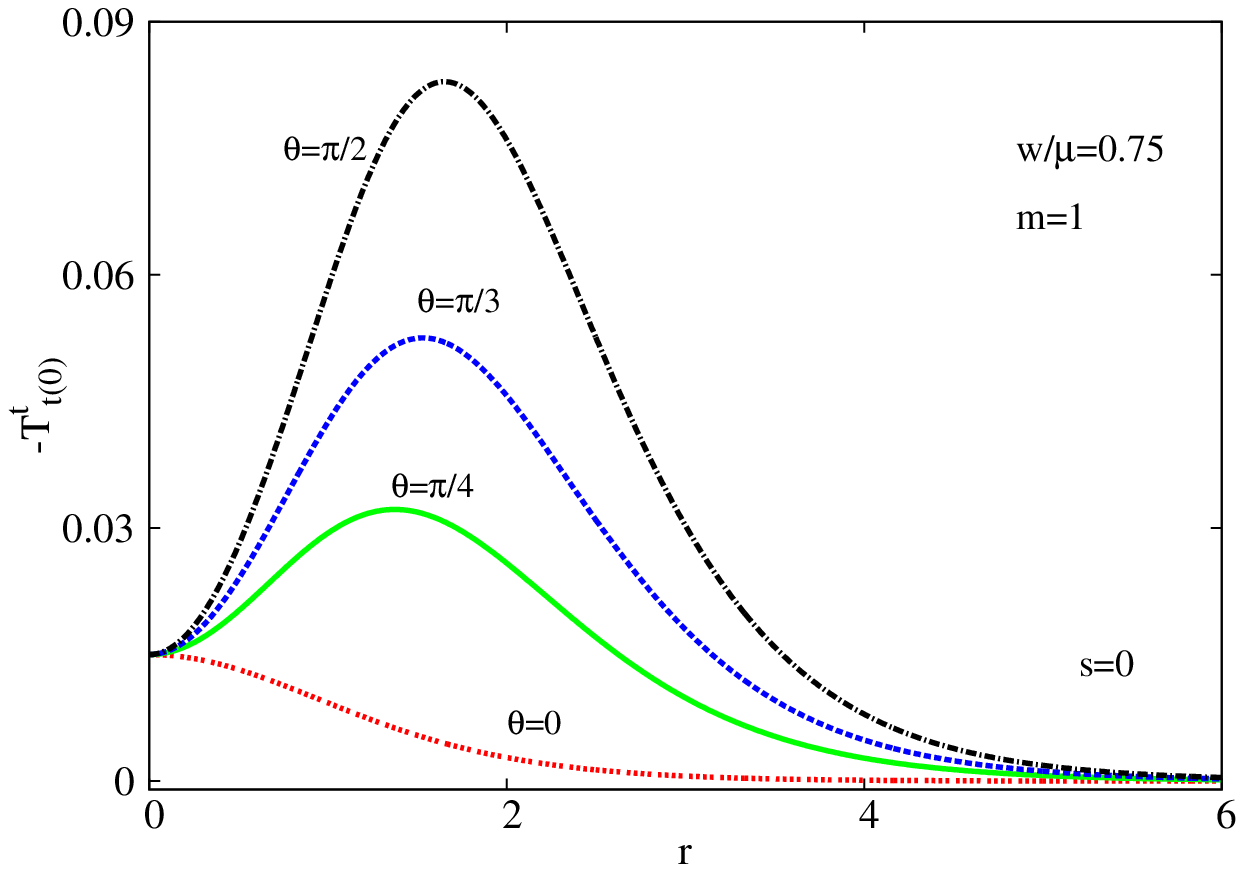}
\includegraphics[width=0.49\textwidth]{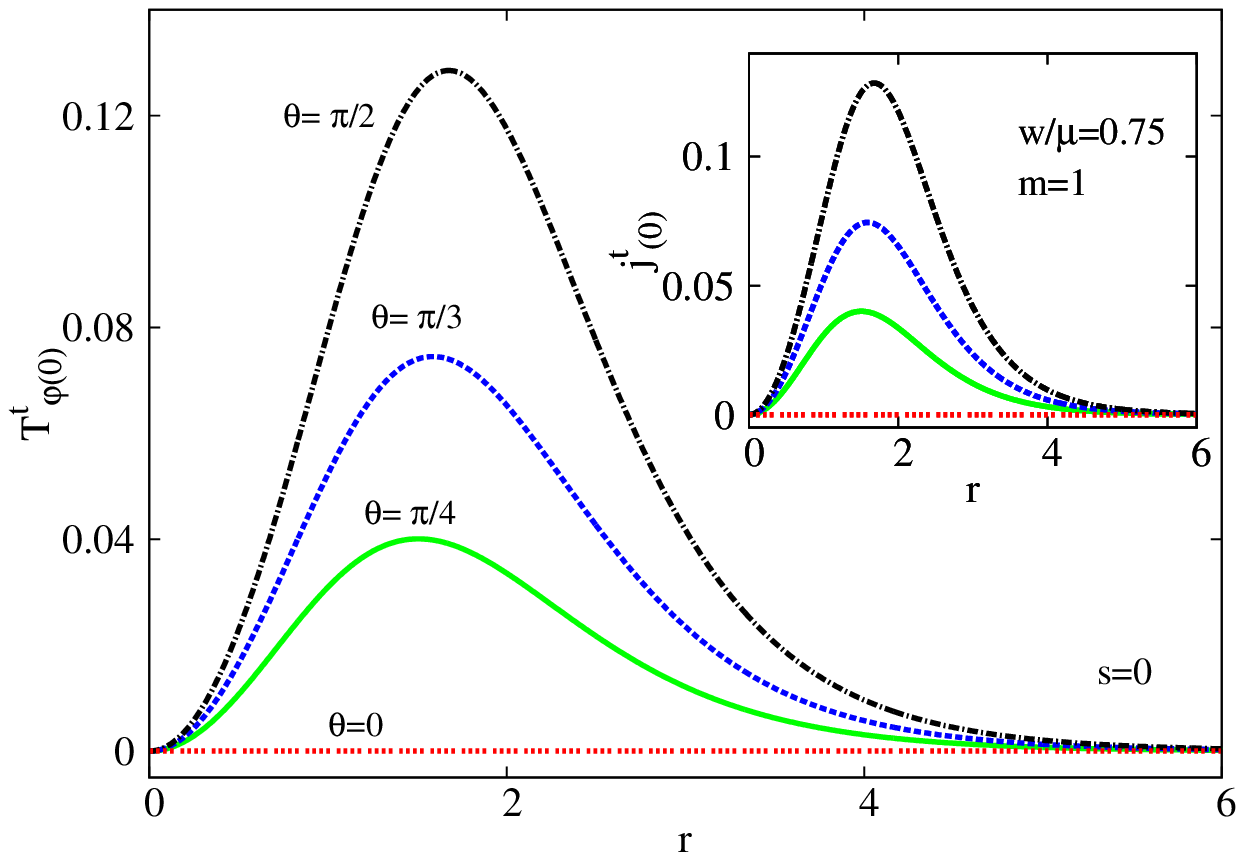}
\includegraphics[width=0.49\textwidth]{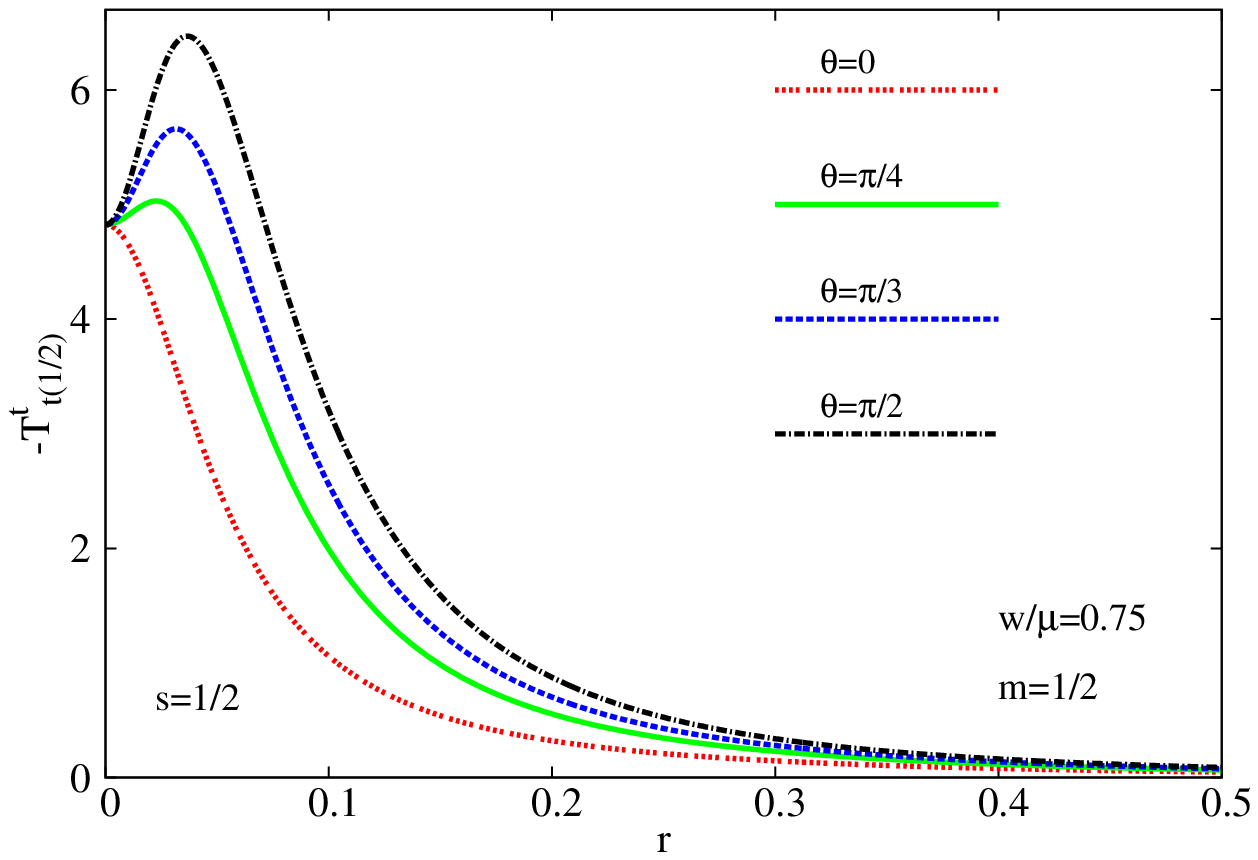}
\includegraphics[width=0.49\textwidth]{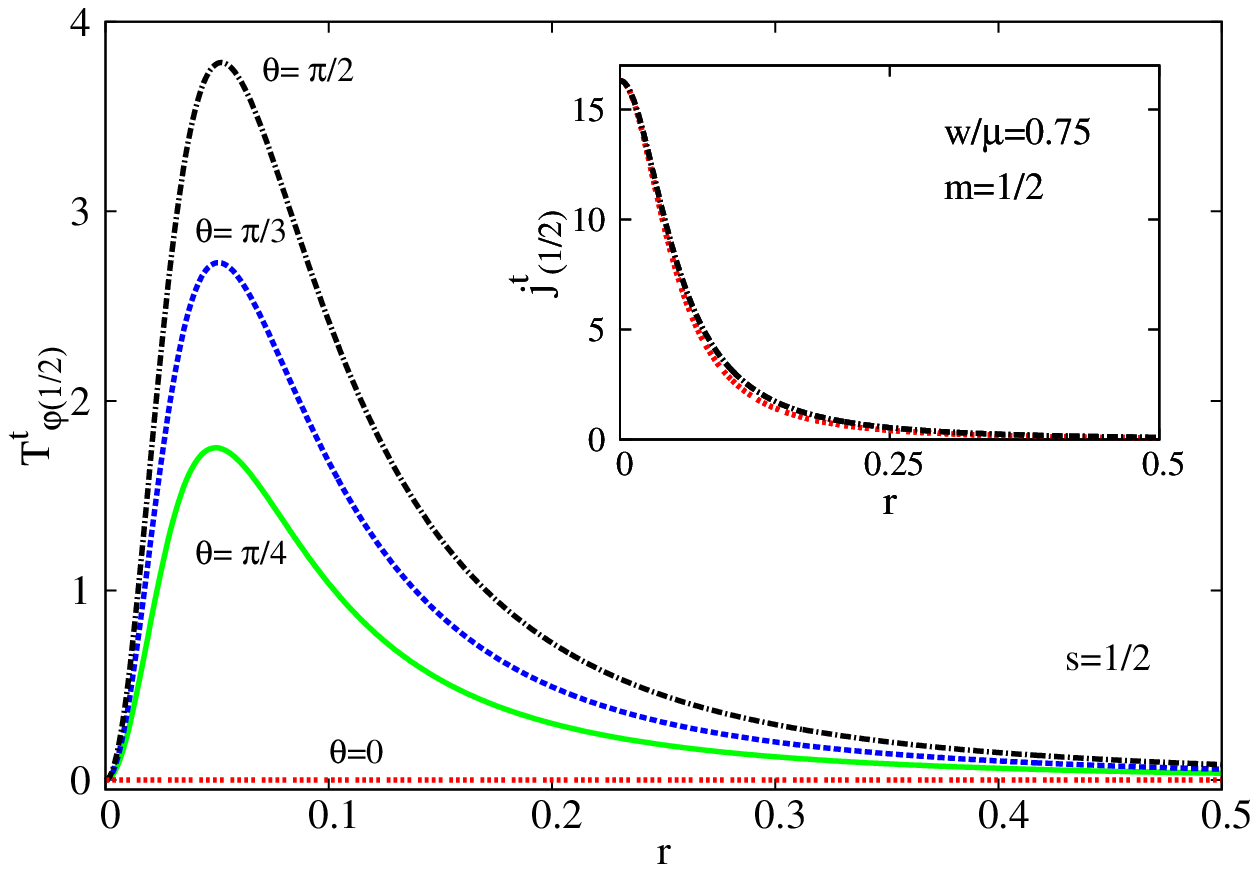}
\includegraphics[width=0.49\textwidth]{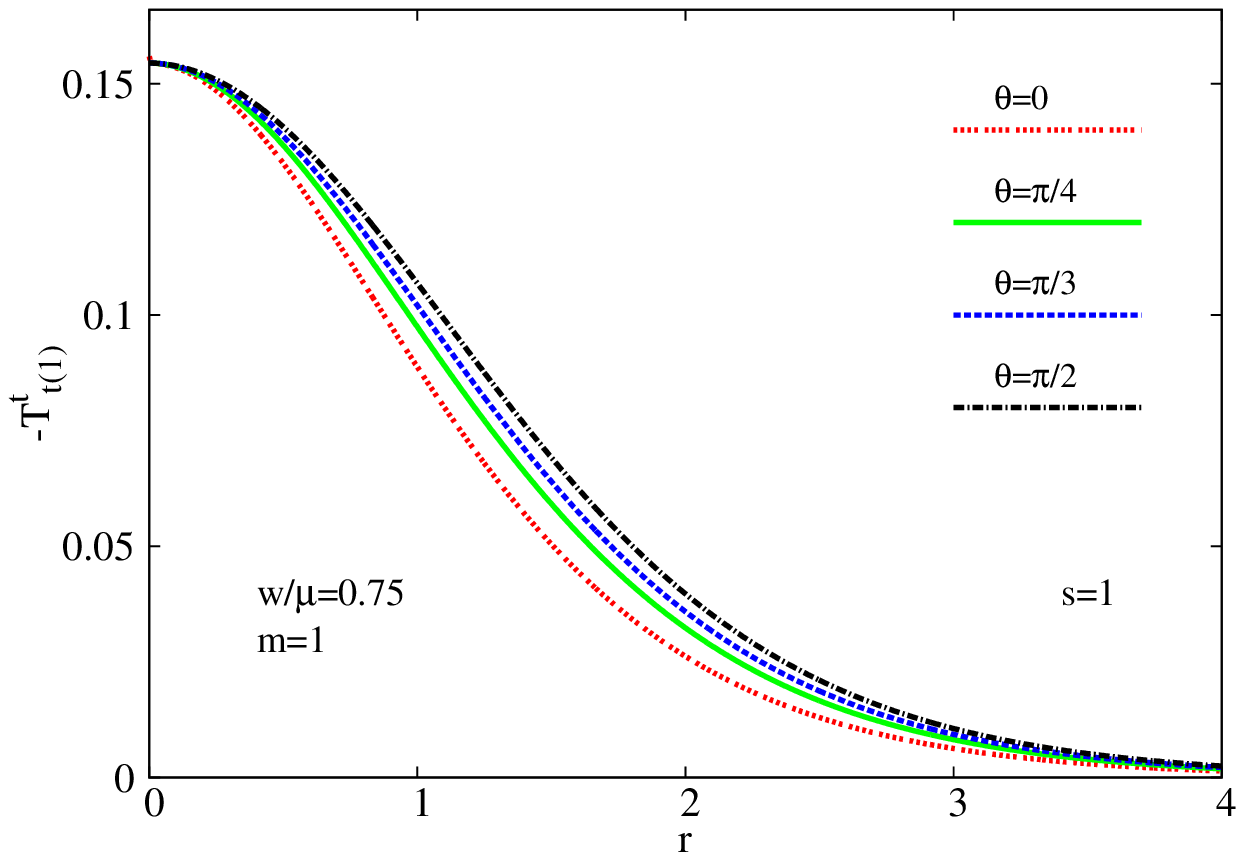}
\includegraphics[width=0.49\textwidth]{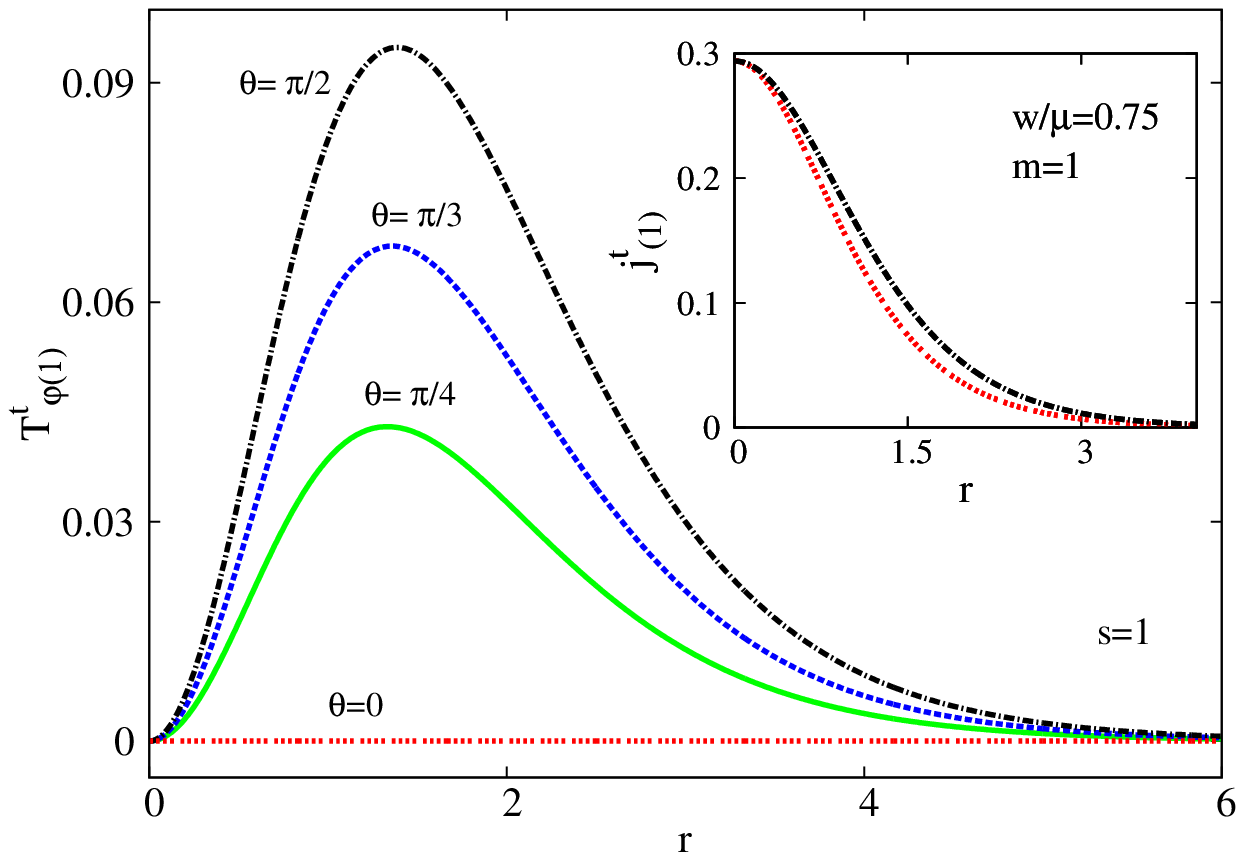}
\caption{\small{The  components $T_t^t$ (left panels) and $T_\varphi^t$ (right panels) of the energy-momentum tensor,
and the 4-current component $j^t$ (right panels inset)
are shown for a fundamental branch solution of the scalar (top panels), Dirac (middle panels) and Proca (bottom panels) model,
all with the same frequency,
$w/\mu=0.75$.
}
}
\label{fig1}
\end{center}
\end{figure}

In Figure \ref{fig1} we display the components $T_t^t$
and $T_\varphi^t$
of the energy-momentum tensor related to the mass-energy and
angular momentum density, together with the temporal component $j^t$ of the current for a
typical  (fundamental branch) solution of each model, all with $w/\mu=0.75$
and the lowest allowed value of $m>0$.
One can see that, unlike for the scalar case, for Dirac and Proca stars, $T_\varphi^t$ and
$j^t$ are not proportional, with the maximum of
  $T_\varphi^t$
 located on the equatorial plane,
while $j^t$ posses an almost spherical shape (the last feature, however, changes for higher $m$).
 A qualitative difference is that both scalar and Dirac stars  possess an intrinsic toroidal shape in what concerns their energy distribution; for the Proca case, however, this distribution is almost spherical.
Another qualitative difference is that for the scalar stars $j^t=0$ on the symmetry axis.
 %

\begin{figure}[h!]
\begin{center}
\includegraphics[width=0.49\textwidth]{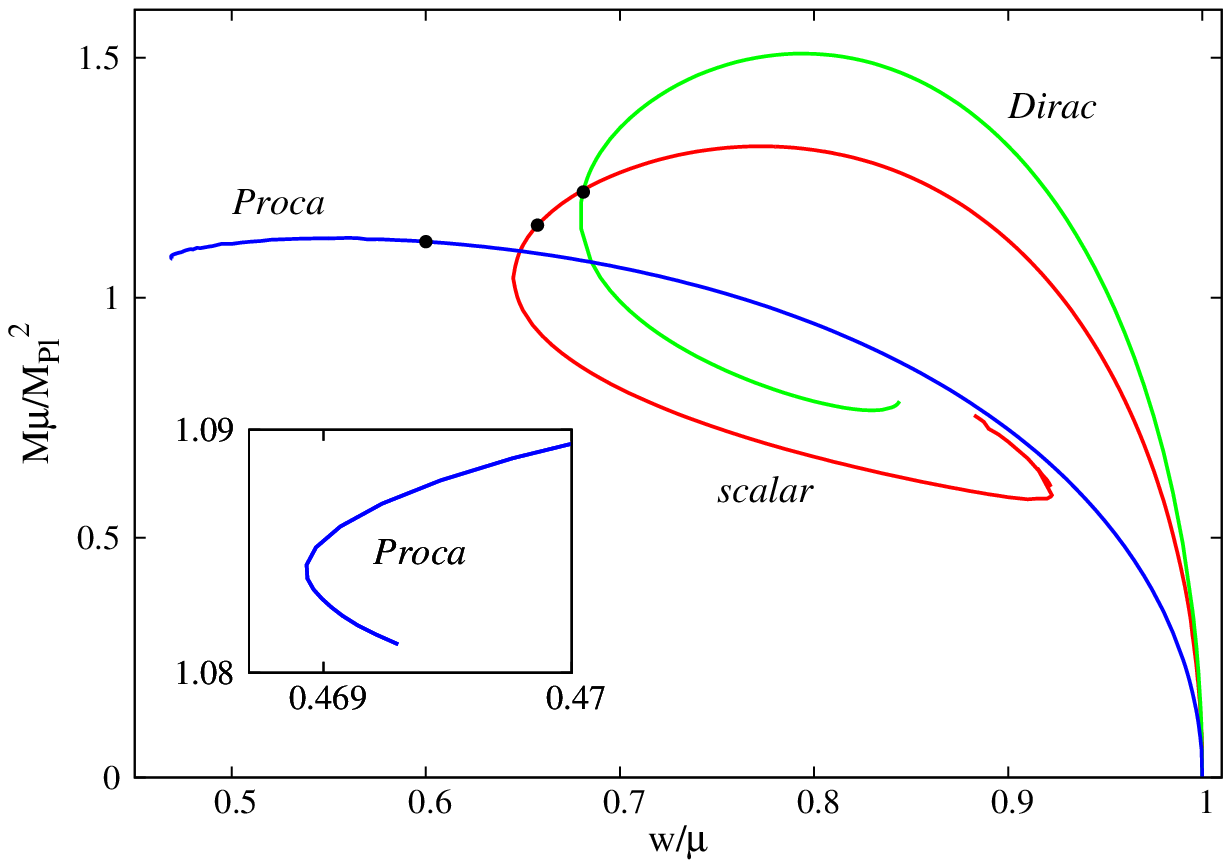}
\includegraphics[width=0.49\textwidth]{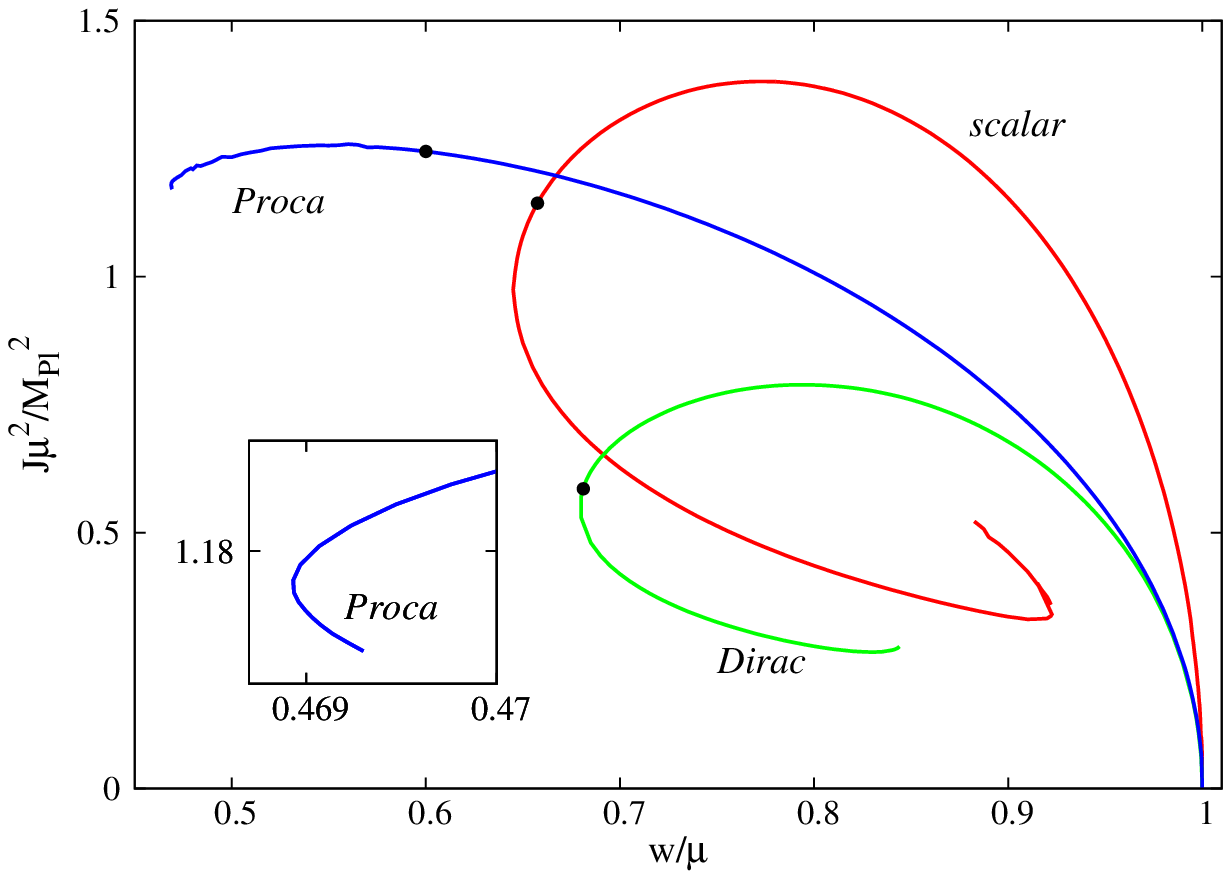}
\caption{\small{The ADM mass $M$ (left panel) and the
angular momentum $J$ (right panel) $vs.$ field frequency $w$
for the  scalar (red line), vector (blue line) and spinor (green line) models.
In each case the dot marks the particular solutions where an ergoregion first occurs, when moving from the maximal frequency $w/\mu=1$ towards the centre of the spiral. The inset provides a zoom on the backbending of the curves, for the Proca case.
}
}
\label{fig2}
\end{center}
\end{figure}

 As seen in Figure \ref{fig2},
in all three cases,
when considering
a mass $M$/angular momentum $J$/Noether charge $Q$, $vs.$  frequency  $w$, diagram,
the domain of existence of the solutions
corresponds to a smooth curve.
This curve
starts from $M=0$ ($J=0$) for $w=\mu$, in which limit the fields becomes very diluted and the solution trivialises.
  At some intermediate frequency, a maximal mass (angular momentum) is attained.
    The parameters of these particular solutions are given in the  2$^{nd}$- 4$^{th}$ columns of Table 1.
    As can be seen there,   the behaviour is not monotonic with spin.
 In each case there is also a minimal frequency, below which no solutions are found.
 The  minimal frequencies and the corresponding $M,J$ are shown in the 5$^{th}$-7$^{th}$ columns of Table~1.
 After reaching the minimal frequency, the spiral backbends into a second branch.
For the scalar and Dirac fields we were able to obtain further backbendings and branches.
For a Proca field,  however,
 we have not been able to construct these secondary branches.
For any value of $s$,
we conjecture that, similarly to the spherically symmetric case,
 the $M(w)$ (and $Q(w)$) curves describe
spirals which approach, at their center, a critical singular solution.

As expected, in all three cases, rotating solutions in the strong gravity region possess an ergo-region of toroidal shape~\cite{Herdeiro:2014jaa}.
The position of the critical solutions for which the ergo-region emerges is shown with a dot in Figure \ref{fig2}.
All remaining solutions, starting from that particular configuration up to the putative solution at the centre of the spiral,
have an $S^1\times S^1$ ergo-surface.

\begin{table}[h!]
{\small \begin{center}
\begin{tabular}{c|||c|c|c||c|c|c||c|c}
\hline
& \ \               $M^{\rm max}$ & \ \ $J^{\rm max}$ & \ \ $w(M^{\rm max},J^{\rm max})$ & \ \  $w^{\rm min}$ & \ \  $M(w^{\rm min})$  & \ \  $J(w^{\rm min})$  & $M=Q$ & \ \ $w^{\rm crossing}$\\
\hline
scalar \ \  & \ \   1.315 \ \  & \ \   1.381 \ \  & \ \     0.775 \ \ &  \ \  0.645 \ \ & \ \ 1.041  \ \ & \ \ 0.975 \ \ & \ \ 1.166 & \ \ 0.661 \\
Dirac  \ \  & \ \   1.509 \ \  & \ \   0.789 \ \  & \ \     0.795 \ \  & \ \  0.680 \ \ & \ \ 1.198  \ \ & \ \ 0.569 \ \ & \ \ 1.303 & \ \ 0.692  \\
Proca  \ \  & \ \   1.125 \ \  & \ \   1.259 \ \  & \ \     0.562 \ \  & \ \  0.469 \ \ & \ \ 1.086  \ \ & \ \ 1.180 \ \ & \ \ - & \ \ - \\
 \hline
\end{tabular}\\
\bigskip
 \caption{1$^{st}$ column: the three different models. 2$^{nd}$, 3$^{rd}$ and 4$^{th}$ columns: mass, angular momentum and frequency of the solution with maximal mass and angular momentum;
5$^{th}$, 6$^{th}$ and 7$^{th}$ columns: frequency, mass and angular momentum of the minimal frequency solution - first backbending in the   diagrams of Fig.~\ref{fig1}; 
 8$^{th}$-9$^{th}$ columns: mass/Noether charge and frequency of the solution with equal ADM mass and Noether charge (the data for Proca stars
is missing in this case).
 All quantities are presented in units of $\mu$, $G$. }
\end{center}}
\end{table}

Although a detailed stability analysis of this solutions is technically challenging and beyond the scope of this paper, some simple observations can be done based on energetic arguments.  The Noether charge measures  the particle number. If this quantity multiplied by the field mass $\mu$ is smaller than the ADM mass $M$,
then the solution has excess, rather than binding, energy and it should be unstable against fission.
 In all three cases, close to the maximal frequency,
$w=\mu$ the solutions are stable under this criterion: there is binding energy, a necessary, albeit not sufficient,  condition for stability.
 For scalar and spinor fields, we have found that at some point,
 the Noether charge and ADM mass curves cross and $M$ becomes larger than $Q$ corresponding to solutions with excess energy and hence unstable.
The corresponding parameters of these
particular solutions
 are given in the 8$^{th}$-9$^{th}$  columns of Table 1.
 A similar picture should exist for Proca stars as well, but so far we have not been able to construct the corresponding solutions.

We emphasise that solutions with binding energy may, nonetheless, be perturbatively unstable.
This has been clarified so far only for spherically symmetric configurations -- 
see Refs. \cite{Gleiser:1988ih,Lee:1988av} for $s=0$,
Refs. \cite{Brito:2015pxa,Sanchis-Gual:2017bhw}
for $s=1$
and Ref. \cite{Finster:1998ws} for $s=1/2$.

\subsection{Bosonic $vs.$ fermionic nature}

\begin{figure}[h!]
\begin{center}
\includegraphics[width=0.495\textwidth]{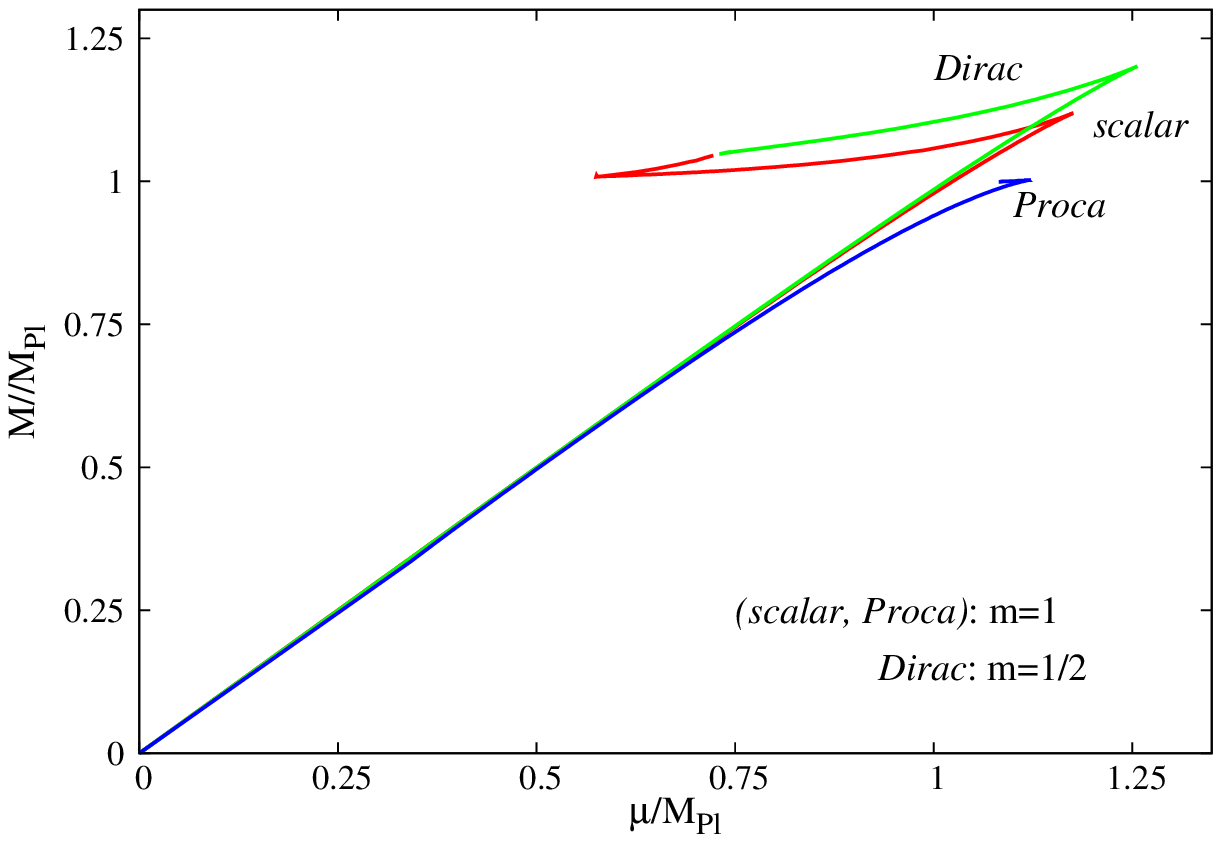}
\includegraphics[width=0.495\textwidth]{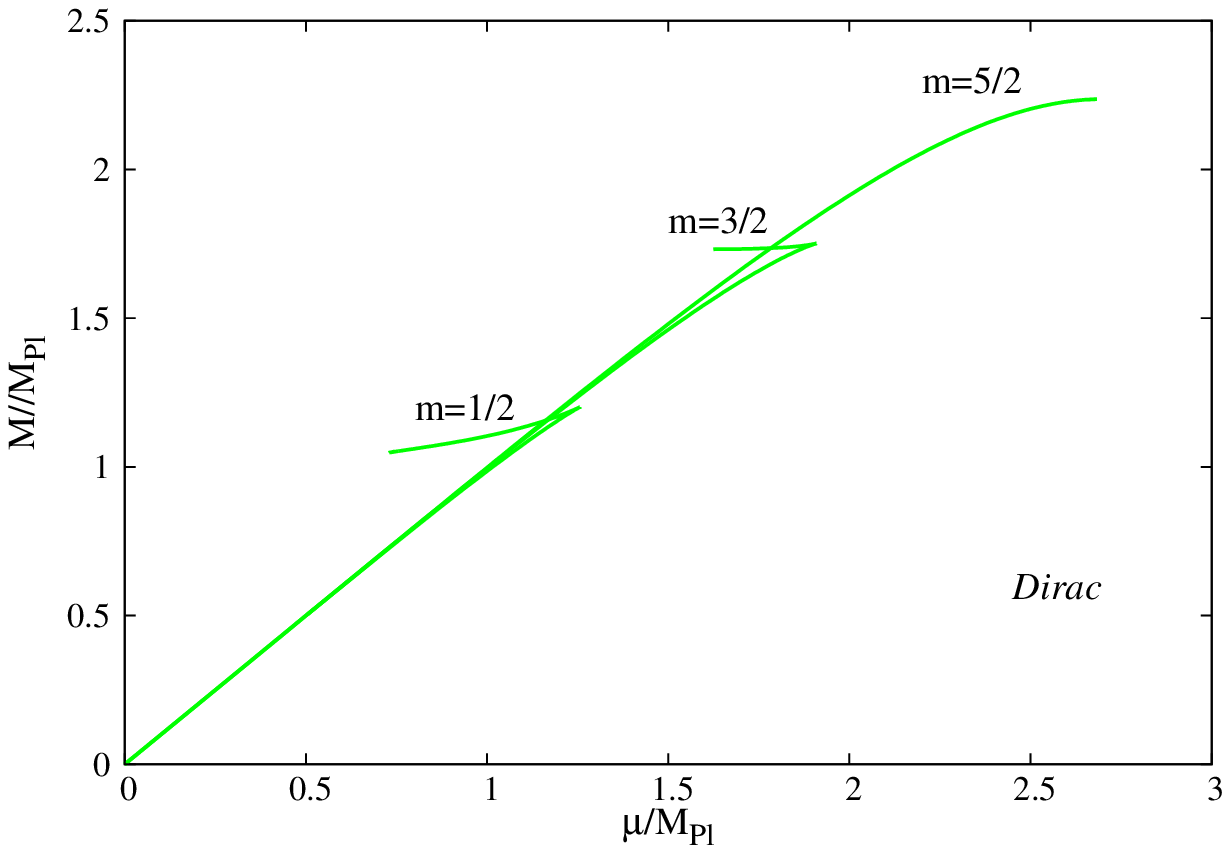}
\caption{\small{Consequences of the single particle condition $Q=1$. (Left panel) ADM mass $vs.$ scalar field mass,
in Planck units, for the three families of stars.
(Right panel) Same for the first three states of the Dirac field ($m=1/2,3/2$ and $m=5/2$).
}}
\label{fig4}
\end{center}
\end{figure}

What if one tries to go beyond the classical field theory analysis and impose the quantum nature of fermions, which demands $Q=1$ for Dirac stars? This condition can also be imposed for scalar and Proca stars, although in those cases it is not a mandatory requirement.
Then, as discussed in~\cite{Herdeiro:2017fhv}, the spiral in Figure~\ref{fig2}
is not a sequence of solutions with constant $\mu$ and varying $Q$ -- recall that here $J=mQ$ --; rather,
 it is a sequence with constant $Q$ and varying $\mu$.
Thus, since $\mu$ is a parameter in the action,  it represents a sequence of solutions of different models.
Consequently, there cannot be a difference of orders of
magnitude between $M$, the physical mass of the star, and $\mu$,  the mass of the field. They should be of the same order of magnitude, unlike the macroscopic quantum states that may occur in the bosonic case.
This is illustrated in Figure~\ref{fig4} (left panel),
where we plot the same data as in Figure~\ref{fig1}
but imposing the single particle condition.

Considering the stars as one particle microscopic classical configurations, the mass of the field $\mu$ becomes bounded, and, for fundamental states, never exceeds, $\sim M_{Pl}$.  Thus,  for these single particle configurations, the particle's size (measured by its Compton wavelength) cannot be smaller than $\sim$ Planck length. This upper $\mu$ bound can be pushed further up by considering configurations with higher values of $m$,
making these configurations increasingly trans-Planckian.
The corresponding masses for the Dirac model with $m=1/2,3/2$ and $5/2$ are shown in Figure~\ref{fig4} (right panel).

\section{Further remarks}

The main purpose of this work was to provide a comparative analysis
of three different types of spinning solitonic solutions of General Relativity coupled with
matter fields of spin
$0$, $1$ and $1/2$, respectively.
In particular, the Einstein-Dirac spinning configurations are reported here for the first time. In all cases there is a harmonic time
dependence in the fields (with a frequency $w$), together with a confining mechanism, as provided
by a mass $\mu$ of the elementary quanta of the field.

Our results confirm that, when considered as classical field theory solutions, the stars share the same universal pattern,
insensitive to the fermionic/bosonic nature of the fields.
That is, when ignoring 
Pauli's exclusion principle, the (field frequency-mass/Nother charge)-diagram of the solutions looks
similar for both bosonic and fermionic stars.\footnote{As discussed in \cite{Blazquez-Salcedo:2019qrz}, this holds also for  the higher dimensional  spherical stars.}
This generalizes the results in
\cite{Herdeiro:2017fhv}
for  spherically symmetric configurations.
Introducing spin, another universal feature is the relation
(\ref{JQ}),
$i.e.$ the angular momentum and the particle number are always proportional
(although the situation is more subtle for Proca and Dirac fields). We conjecture that similar configurations may exist for \textit{any} spin,
given a consistent matter model minimally coupled to GR, likely with similar properties.
In particular, this should hold for $s=3/2$: Rarita-Schwinger  stars should exist,
which, for a single field, should also satisfy  relation (\ref{JQ}).

On the other hand, if one imposes that the configuration describes a single particle, which is a consequence of the quantum nature of fermions, one finds that for each field mass there is a discrete set of
 states, up to a maximal field mass.

\medskip
As noticed in \cite{Herdeiro:2017fhv} for the
 spherically symmetric case,
 the observed similarities between bosonic and fermionic
solitons remain in the absence of gravity as long as appropriate self-interactions of the matter fields are allowed.
For the matter fields in this work,
spinning flat space solitons are known for $s=0$ only
\cite{Volkov:2002aj,Kleihaus:2005me}, but should exist for $s=1/2,1$ as well. Moreover, one can show that
the relation (\ref{JQ}) is still satisfied.
A preliminary numerical analysis indicates the existence of
spinning flat space Dirac solitons,
which generalise the solutions in \cite{Soler:1970xp}
 for a $single$ spinor with a quartic self-interaction.

An important difference between bosonic and fermionic solutions is the following.
Scalar or Proca stars can be in equilibrium with a black hole horizon at their centre, if both are rotating synchronously, leading to black holes with scalar or Proca hair~\cite{Herdeiro:2014goa,Herdeiro:2016tmi}.
This does not seem to be the case for a Dirac star. Conventional wisdom may attempt to relate this putative impossibility to  the absence of
superradiance for a fermionic field on the Kerr background \cite{Brito:2015oca}.
However, spinning black holes
with scalar hair exist even in the absence of the superradiant instability, the hair being intrinsically non-linear
\cite{Brihaye:2014nba,Herdeiro:2015kha}.
Therefore one cannot rule out, based on this association, that Dirac stars could
allow for black hole generalisations.
A more convincing obstacle is provided by the following argument.
When assuming the existence of a power series expansion of the Einstein-matter field equations
in the vicinity the event horizon,\footnote{Here it is convenient to consider
again spheroidal coordinates together with a non-extremal horizon.}
the case of a Dirac field appears to be special.
On the one hand, for a bosonic field ($s=0,1$),
the synchronization condition $w=m\Omega_H$
(with $\Omega_H$ the event horizon velocity)
occurs naturally, allowing for non-zero values of the matter fields at the horizon
together with finite values for relevant quantities ($e.g.$ $j^t$).
As a result, a consistent local, non-trivial solution exists, in term of the values taken at the horizon.
On the other hand, this is not the case for a Dirac field, where  the condition
$w=m\Omega_H$ (which still occurs naturally) is not enough to assure regularity at the horizon.
It turns out that the spinor components are forced to vanish there order by order, yielding only the trivial solution.
Despite this suggestive argument,  a rigorous proof of the impossibility of endowing a Kerr black hole with synchronous Dirac hair is still lacking.

Beyond the matter contents discussed in this work,
it is worth mentioning the case of $SU(2)$ Yang-Mills fields.
While this nonlinear model possesses no flat spacetime solitons
\cite{Coleman:1977hd}, the coupling to
gravity allows for particle-like solutions \cite{Bartnik:1988am}.
Spinning
generalisations of these solutions, however, do not exist \cite{VanderBij:2001nm}, a rather unique situation
amongst field theory models.
Nonetheless, spinning Einstein-Yang-Mills configurations
are found when adding a rotating horizon
at the center of a static soliton \cite{Kleihaus:2000kg}.

\section*{Acknowledgements}
%
This work is supported by the Fundacao para a Ci\^encia e a Tecnologia (FCT)
project UID/MAT/04106/2019 (CIDMA), by CENTRA (FCT) strategic project UID/FIS/00099/2013, by national funds (OE), through FCT, I.P., in the scope of the framework contract foreseen in the numbers 4, 5 and 6
of the article 23, of the Decree-Law 57/2016, of August 29,
changed by Law 57/2017, of July 19. We acknowledge support  from the project PTDC/FIS-OUT/28407/2017.
This work has further been supported by  the  European  Union's  Horizon  2020  research  and  innovation
(RISE) programmes H2020-MSCA-RISE-2015
Grant No.~StronGrHEP-690904 and H2020-MSCA-RISE-2017 Grant No.~FunFiCO-777740.
E.R. and Ya.S. gratefully acknowledge the support of the Alexander von Humboldt Foundation.
Ya.S. acknowledges the support from the Ministry of Education and Science of Russian Federation, project No
3.1386.2017. The authors would like to acknowledge networking support by the
COST Action CA16104.

 \begin{small}
 
 \end{small}

 \end{document}